\documentclass[aps,prl,twocolumn,showpacs,superscriptaddress,floatfix]{revtex4-1}

\usepackage{graphicx}
\usepackage{dcolumn}
\usepackage{bm}
\usepackage{amssymb}
\usepackage{microtype}
\usepackage{xfrac}
\usepackage{gensymb}
\usepackage{array}
\usepackage{color}

\begin{document}

\title{Phase Competition in the Palmer-Chalker XY Pyrochlore Er$_2$Pt$_2$O$_7$}

\author{A.~M.~Hallas}
\affiliation{Department of Physics and Astronomy, McMaster University, Hamilton, ON, L8S 4M1, Canada}

\author{J.~Gaudet}
\affiliation{Department of Physics and Astronomy, McMaster University, Hamilton, ON, L8S 4M1, Canada}

\author{N.~P.~Butch}
\affiliation{Center for Neutron Research, National Institute of Standards and Technology, MS 6100 Gaithersburg, Maryland 20899, USA}

\author{Guangyong~Xu}
\affiliation{Center for Neutron Research, National Institute of Standards and Technology, MS 6100 Gaithersburg, Maryland 20899, USA}

\author{M.~Tachibana}
\affiliation{National Institute for Materials Science, 1-1 Namiki, Tsukuba 305-0044, Ibaraki, Japan}

\author{C.~R.~Wiebe}
\affiliation{Department of Physics and Astronomy, McMaster University, Hamilton, ON, L8S 4M1, Canada}
\affiliation{Department of Chemistry, University of Winnipeg, Winnipeg, MB, R3B 2E9 Canada}
\affiliation{Canadian Institute for Advanced Research, 180 Dundas St. W., Toronto, ON, M5G 1Z7, Canada}

\author{G.~M.~Luke}
\affiliation{Department of Physics and Astronomy, McMaster University, Hamilton, ON, L8S 4M1, Canada}
\affiliation{Canadian Institute for Advanced Research, 180 Dundas St. W., Toronto, ON, M5G 1Z7, Canada}

\author{B.~D.~Gaulin}
\affiliation{Department of Physics and Astronomy, McMaster University, Hamilton, ON, L8S 4M1, Canada}
\affiliation{Canadian Institute for Advanced Research, 180 Dundas St. W., Toronto, ON, M5G 1Z7, Canada}
\affiliation{Brockhouse Institute for Materials Research, Hamilton, ON L8S 4M1 Canada}

\date{\today}

\begin{abstract} 
We report neutron scattering measurements on Er$_2$Pt$_2$O$_7$, a new addition to the XY family of frustrated pyrochlore magnets. Symmetry analysis of our elastic scattering data shows that Er$_2$Pt$_2$O$_7$ is the first XY pyrochlore to order into the $k=0$, $\Gamma_7$ magnetic structure (the Palmer-Chalker state), at $T_N = 0.38$~K. This contrasts with its sister XY pyrochlore antiferromagnets Er$_2$Ti$_2$O$_7$ and Er$_2$Ge$_2$O$_7$, both of which order into $\Gamma_5$ magnetic structures at much higher temperatures, $T_N=1.2$~K and $1.4$~K, respectively. In this temperature range, the magnetic heat capacity of Er$_2$Pt$_2$O$_7$ contains a broad anomaly centered at $T^*=1.5$~K. Our inelastic neutron scattering measurements reveal that this broad heat capacity anomaly sets the temperature scale for strong short-range spin fluctuations. Below $T_N = 0.38$~K, Er$_2$Pt$_2$O$_7$ displays a gapped spin wave spectrum with an intense, flat band of excitations at lower energy and a weak, diffusive band of excitations at higher energy. The flat band is well-described by classical spin wave calculations, but these calculations also predict sharp dispersive branches at higher energy, a striking discrepancy with the experimental data. This, in concert with the strong suppression of $T_N$, is attributable to enhanced quantum fluctuations due to phase competition between the $\Gamma_7$ and $\Gamma_5$ states that border each other within a classically predicted phase diagram.
\end{abstract}

\maketitle

The low temperature magnetism of the rare-earth pyrochlore oxides, $R_2$$B_2$O$_7$, has become synonymous with complexity and exotic ground states. Both of these are natural consequences of magnetism on the pyrochlore lattice, which is comprised of two site-ordered networks of corner-sharing tetrahedra. This is the canonical three-dimensional crystalline architecture for geometric magnetic frustration, in which competing interactions can preclude or hinder the formation of a classically ordered state. The diversity in the phenomenology of the rare-earth pyrochlores is attributable to the different anisotropies and interactions exhibited by the rare-earth ions that can occupy its magnetic sublattice, which conspire to produce a veritable zoo of magnetic behaviors~\cite{gardner2010magnetic}.

A particularly interesting sub-group of the rare-earth pyrochlores are those that exhibit XY spin anisotropy~\cite{hallas2017review}, which is obtained when the rare-earth site is occupied by either erbium (Er) or ytterbium (Yb). This XY label is garnered on the basis of their crystal electric field phenomenology, where in both cases, the ground state is an isolated doublet protected by Kramers' theorem, allowing an effective $S=\sfrac{1}{2}$ description~\cite{guitteny2013palmer,gaudet2015neutron,gaudet2017effect}. The anisotropic exchange Hamiltonian, with a form determined by the symmetry of the crystal lattice, provides an appropriate starting point for understanding the ground states of many XY pyrochlores~\cite{ross2011quantum,savary2012order,yan2017theory}. Within the nearest neighbor version of this model, certain sets of exchange parameters can give rise to exotic states such as quantum spin ice~\cite{ross2011quantum,gingras2014quantum} or various spin liquids~\cite{canals1998pyrochlore,benton2016spin}, while other sets of exchange parameters are predicted to stabilize classically ordered states~\cite{yan2017theory}. 

\begin{figure}[h]
\linespread{1}
\par
\includegraphics[width=3.3in]{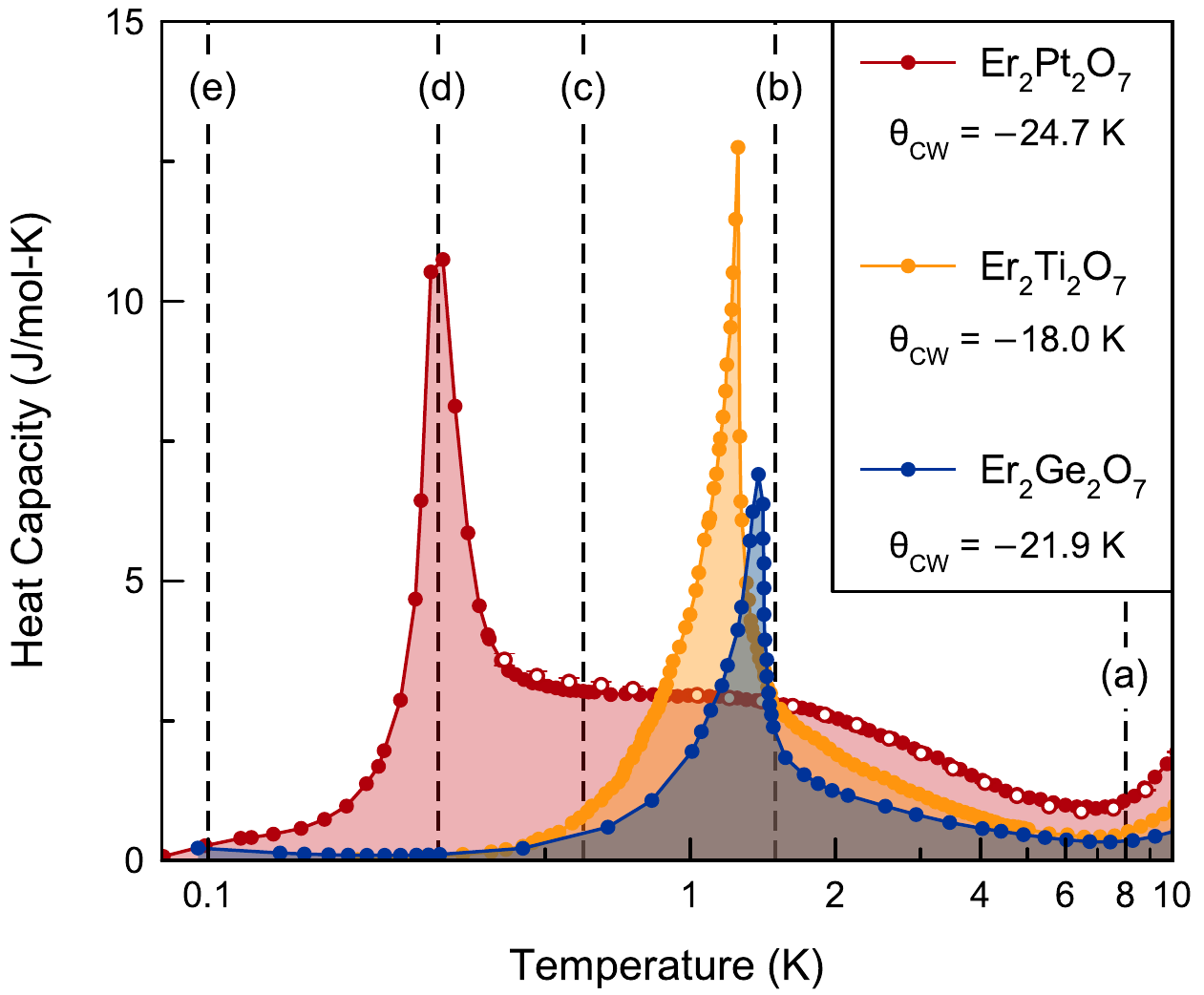}
\par
\caption{Low temperature heat capacity of the three sister XY antiferromagnets: Er$_2$Ge$_2$O$_7$~\cite{dun2015antiferromagnetic}, Er$_2$Ti$_2$O$_7$~\cite{de2012magnetic}, and Er$_2$Pt$_2$O$_7$, filled in circles from~\cite{cai2016high} and open circles from our study. The latter two samples magnetically order with $T_N=1.4$~K and 1.2~K, respectively. In this temperature range, Er$_2$Pt$_2$O$_7$ exhibits a broad heat capacity anomaly centered at $T^*=1.5$~K with a strongly suppressed $T_N$. The vertical dashed lines indicate the temperatures that correspond with the inelastic scattering spectra that are presented in Figure~\ref{Inelastic}.}
\label{HeatCapacity}
\end{figure}

The phase diagram that encompasses the region of parameter space believed to be relevant to the XY pyrochlores contains four distinct $k=0$ ordered states~\cite{wong2013ground,yan2017theory}: $\psi_2$ non-coplanar antiferromagnet, $\psi_3$ coplanar antiferromagnet, $\Gamma_7$ antiferromagnet (the Palmer-Chalker state~\cite{palmer2000order}), and $\Gamma_9$ splayed ferromagnet. Of these states, all but $\Gamma_7$ have been experimentally observed in the XY pyrochlores. This XY family is made up of Yb$_2B_2$O$_7$ and Er$_2B_2$O$_7$ with $B=$~Ge, Ti, and Sn, where: (i) The order-by-disorder candidate Er$_2$Ti$_2$O$_7$ orders into $\psi_2$ \cite{poole2007magnetic} (ii) Er$_2$Ge$_2$O$_7$ and Yb$_2$Ge$_2$O$_7$ have as-of-yet unidentified ordered states within $\Gamma_5$ ($\psi_2$ or $\psi_3$) \cite{dun2015antiferromagnetic,hallas2016xy} and (iii) Both Yb$_2$Sn$_2$O$_7$ \cite{yaouanc2013dynamical,lago2014glassy} and some samples of Yb$_2$Ti$_2$O$_7$ \cite{yasui2003ferromagnetic,chang2012higgs,gaudet2016gapless,yaouanc2016novel,scheie2017lobed} order into the $\Gamma_9$ splayed ferromagnetic state. This ensemble of magnetic ground states supports the picture of a rich phase space. 

In this Letter, we present a comprehensive neutron scattering study of Er$_2$Pt$_2$O$_7$, a recent addition to the XY family of pyrochlores~\cite{cai2016high,gaudet2017effect}. Through magnetic symmetry analysis, we find that Er$_2$Pt$_2$O$_7$ orders into the $\Gamma_7$ Palmer-Chalker state, the first XY pyrochlore shown to possess this ground state. The N\'{e}el ordering temperature, $T_N = 0.38$~K, is a 75\% reduction from those of its closest sister pyrochlore antiferromagnets: Er$_2$Ti$_2$O$_7$ and Er$_2$Ge$_2$O$_7$. Given that the lattice parameter of Er$_2$Pt$_2$O$_7$ differs by less then 0.5\% from its titanate analog, it is surprising that the transition temperature is so substantially reduced. This dramatic reduction in $T_N$ occurs despite minimal structural modifications and a larger Curie-Weiss temperature, as given in Figure~\ref{HeatCapacity}. Our inelastic neutron scattering measurements reveal that strong quasi-elastic spin fluctuations develop in Er$_2$Pt$_2$O$_7$ at a temperature well-above $T_N$, around $T^*=1.5$~K. This is coincident with the $T_N$'s of both Er$_2$Ti$_2$O$_7$ and Er$_2$Ge$_2$O$_7$, and a broad peak in its own magnetic heat capacity, as shown in Figure~\ref{HeatCapacity}. Below $T_N=0.38$~K, Er$_2$Pt$_2$O$_7$'s spin wave spectrum contains a narrow band of low energy spin excitations that are gapped by $0.18\pm0.02$~meV from the elastic position and a diffusive band at higher energy. Spin wave calculations show that the spin excitation spectrum of Er$_2$Pt$_2$O$_7$ should contain dispersive higher energy branches that are strikingly absent from the experimental data. We conclude that the origin of the suppressed $T_N$ and the unusual spin dynamics is strong phase competition between the $\Gamma_7$ and $\Gamma_5$ states. 

\begin{figure}[tbp]
\linespread{1}
\par
\includegraphics[width=3.3in]{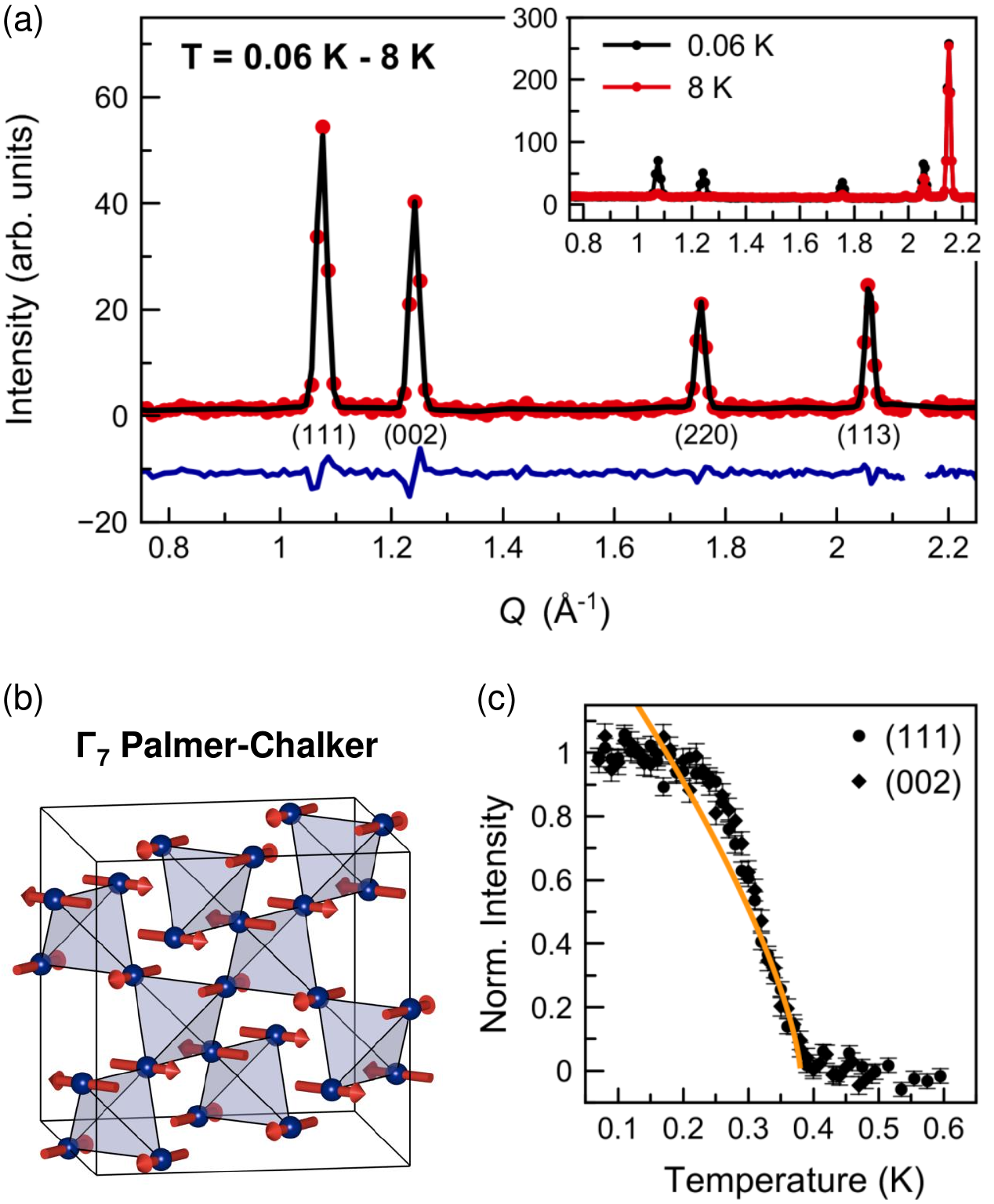}
\par
\caption{Rietveld refinement of Er$_2$Pt$_2$O$_7$ at 0.06~K, where the magnetic scattering has been isolated by subtracting the 8~K data set. The data (red points) is refined against the $\Gamma_7$ magnetic structure, the resulting fit is given by the black curve and the residual is given by the blue curve. The inset shows the unsubtracted elastic scattering at 0.06~K and 8~K. (b) The spin configuration of Er$_2$Pt$_2$O$_7$ in its Palmer-Chalker ($\Gamma_7$) ground state. (c) The intensity of the (111) and (002) magnetic Bragg peaks as a function of temperature normalized by the average high and low temperature values. A power law fit, given by the yellow curve, gives a critical exponent of $\beta = 0.35 \pm 0.03$.}
\label{Refinements}
\end{figure}

Er$_2$Pt$_2$O$_7$ can be synthesized in the cubic $Fd\bar{3}m$ pyrochlore structure, in powder form only, using high-pressure techniques. We investigated the low temperature magnetic state of our 1.2 gram sample of Er$_2$Pt$_2$O$_7$ using both elastic and inelastic neutron scattering techniques. Elastic measurements were performed on the cold neutron triple-axis spectrometer SPINS and time-of-flight inelastic measurements were performed on the Disc Chopper Spectrometer~\cite{copley2003disk}, both located at the National Institute for Standards and Technology's Center for Neutron Research. Further details of the synthesis and experimental methods can be found in the Supplemental Material.

\begin{figure*}[htbp]
\linespread{1}
\par
\includegraphics[width=7in]{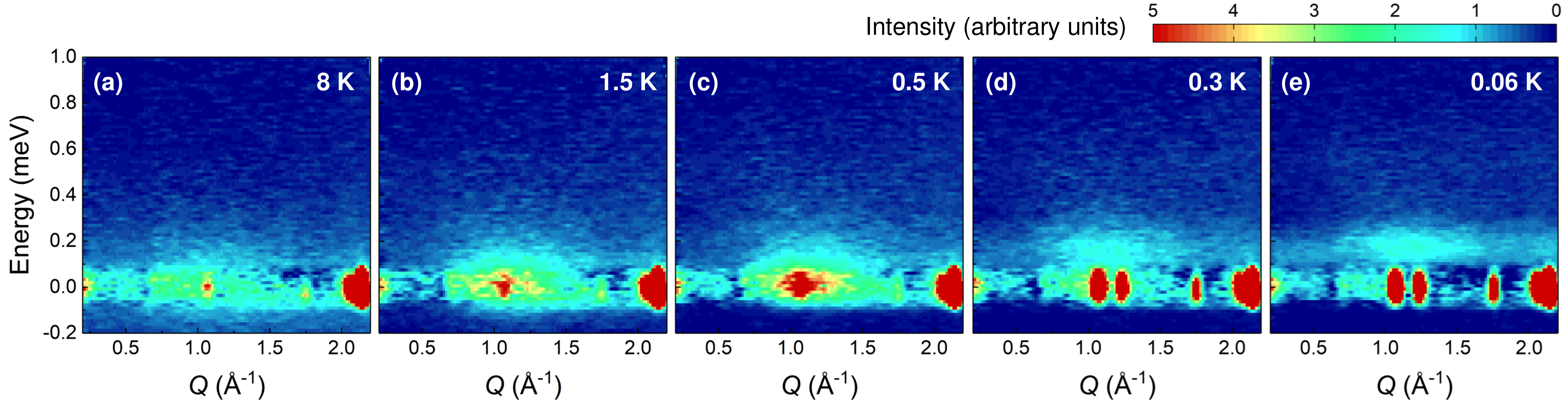}
\par
\caption{The inelastic neutron scattering spectra for Er$_2$Pt$_2$O$_7$ at (a) 8~K, (b) 1.5~K, (c) 0.5~K, (d) 0.3~K, and (e) 0.06~K. Each data set has had an empty sample can background subtracted. At $T^*=1.5$~K, the center of the broad specific heat anomaly, short range correlations are building up at 1.1~\AA$^{-1}$; these correlations grow more intense down to 0.5~K. Below $T_N=0.38$~K, the spectral weight segregates into magnetic Bragg peaks and a gapped spin wave excitation.}
\label{Inelastic}
\end{figure*}

The magnetically ordered state of Er$_2$Pt$_2$O$_7$ can be characterized by the Bragg scattering, which we isolate by integrating over the elastic channel ($\pm 0.05$~meV) in the time-of-flight data. As shown in the inset to Figure~\ref{Refinements}(a), additional Bragg scattering forms upon cooling from 8~K to 0.06~K due to long range magnetic ordering. These magnetic Bragg peaks are resolution limited, corresponding to a minimum correlation length of $132 \pm 9$~\AA. A new Bragg reflection is observed to form on the (002) position, as well as enhanced intensity on the (111), (220), and (113) positions. These magnetic reflections can all be indexed with the propagation vector $k=0$. The possible $k=0$ magnetic structures for Er$^{3+}$ at the $16d$ crystallographic position in the $Fd\bar{3}m$ space group are described by four irreducible representations: $\Gamma_{\text{mag}} = \Gamma_3^1 + \Gamma_5^2 + \Gamma_7^3 + \Gamma_9^6$, where the superscript denotes the number of basis vectors for the given representation, which are labeled $\psi_1, \psi_2, ... \psi_{12}$~\cite{wills2006magnetic}. Both $\Gamma_3$ and $\Gamma_5$ can be immediately ruled out, as the (002) magnetic reflection is symmetry forbidden in both of these representations, while (002) is very intense in our measured pattern. Furthermore, bulk characterization indicates the ordered state of Er$_2$Pt$_2$O$_7$ is antiferromagnetic \cite{cai2016high}, and $\Gamma_9$ is ferromagnetic. Thus, on qualitative grounds alone, one could deduce that Er$_2$Pt$_2$O$_7$ orders into the $\Gamma_7$ irreducible representation.

To definitively determine the ordered state of Er$_2$Pt$_2$O$_7$, we have performed a Rietveld refinement, the result of which is shown in Figure~\ref{Refinements}(a). The magnetic Bragg scattering was isolated by subtracting a high temperature, 8~K, data set from the 0.06~K data set. All structural and instrumental parameters were fixed according to a refinement of the 8~K data set. Thus, the only parameter allowed to vary for the magnetic refinement at 0.06~K is the size of the ordered moment. Magnetic refinements were attempted with each of the $k=0$ representations, and the best agreement, $\chi^2 = 2.22$, was obtained with $\Gamma_7$, validating our earlier qualitative assessment. Fixing the scale of the magnetic scattering according to the structural component allows us to determine the size of the ordered moment, which is 3.4(2) $\mu_{\text{B}}$ at 0.06~K. This ordered moment is approximately 90\% of the total moment that was recently determined for the crystal field ground state of Er$_2$Pt$_2$O$_7$, $\mu_{\text{CEF}} = 3.9~\mu_{\text{B}}$~\cite{gaudet2017effect}. The normalized intensity of the (111) and (002) Bragg peaks as a function of temperature are plotted in Figure~\ref{Refinements}(c). Fitting a narrow temperature range below $T_N=0.38$~K to a power law gives the critical exponent $\beta = 0.35 \pm 0.03$, consistent with conventional 3D XY universality~\cite{campostrini2001critical}.

In the $\Gamma_7$ ordered state of Er$_2$Pt$_2$O$_7$, all spins lie in the plane perpendicular to the local $\langle$111$\rangle$ axes, that connect the vertices of a tetrahedron to its center. The three basis vectors in the $\Gamma_7$ manifold are denoted as $\psi_4$, $\psi_5$, and $\psi_6$. These three basis vectors are connected by cubic symmetry transformations, meaning that they are necessarily degenerate and that an equiprobable distribution of all domains will be present in zero magnetic field. Thus, we can arbitrarily proceed by visualizing $\psi_4$, which is pictured in Figure~\ref{Refinements}(b). On each tetrahedron, there are two pairs of anti-parallel oriented spins, and all spins are aligned parallel to one of the tetrahedron's edges. 

In Figure~\ref{Inelastic} we present the inelastic neutron scattering spectra for Er$_2$Pt$_2$O$_7$ at 8~K, 1.5~K, 0.5~K, 0.3~K and 0.06~K. These temperatures span the range of both specific heat anomalies displayed by Er$_2$Pt$_2$O$_7$ and are indicated by the dashed vertical lines in Figure~\ref{HeatCapacity}. Each of these data sets has had an empty sample can background subtracted from it. We can associate the broad specific heat anomaly at $T^*=1.5$~K to short range quasi-elastic spin fluctuations, giving rise to a diffuse feature centered at 1.1~\AA$^{-1}$. These short range correlations grow more intense upon cooling to 0.5~K. The majority of the diffuse scattering at these temperatures, above $T_N$, is elastic within our 0.09~meV resolution. 

As Er$_2$Pt$_2$O$_7$ is cooled through its N\'{e}el ordering transition at $T_N=0.38$~K the diffuse scattering segregates into sharp magnetic Bragg reflections, a narrow inelastic mode centered near 0.2~meV, and a higher energy broad distribution of spin excitations. The upper broad band of excitations is centered at 0.6~meV as can be seen by integrating over our full $Q$ range, as presented in Figure~\ref{Cuts}(a). However, this upper band of scattering lacks apparent structure as seen in Figure~\ref{Inelastic}(e). The lower band of spin excitations is gapped from the nlastic line by $0.18\pm0.02$~meV. This gap is essentially constant at all wave-vectors, due to the fact that the band itself is so narrow in energy, with a bandwidth of only 0.1~meV. A dispersionless band of excitations, such as this, has been observed in a number of highly frustrated magnetic systems: for example, the ``weathervane mode" predicted for two-dimensional Kagome systems~\cite{chandra1993anisotropic}, and observed in KFe$_3$(OH)$_6$(SO$_4$)$_2$~\cite{matan2006spin}, as well as the singlet-triplet excitations of the frustrated Shastry-Sutherland system, SrCu$_2$(BO$_3$)$_2$~\cite{gaulin2004high}. 

\begin{figure}[tbp]
\linespread{1}
\par
\includegraphics[width=3.5in]{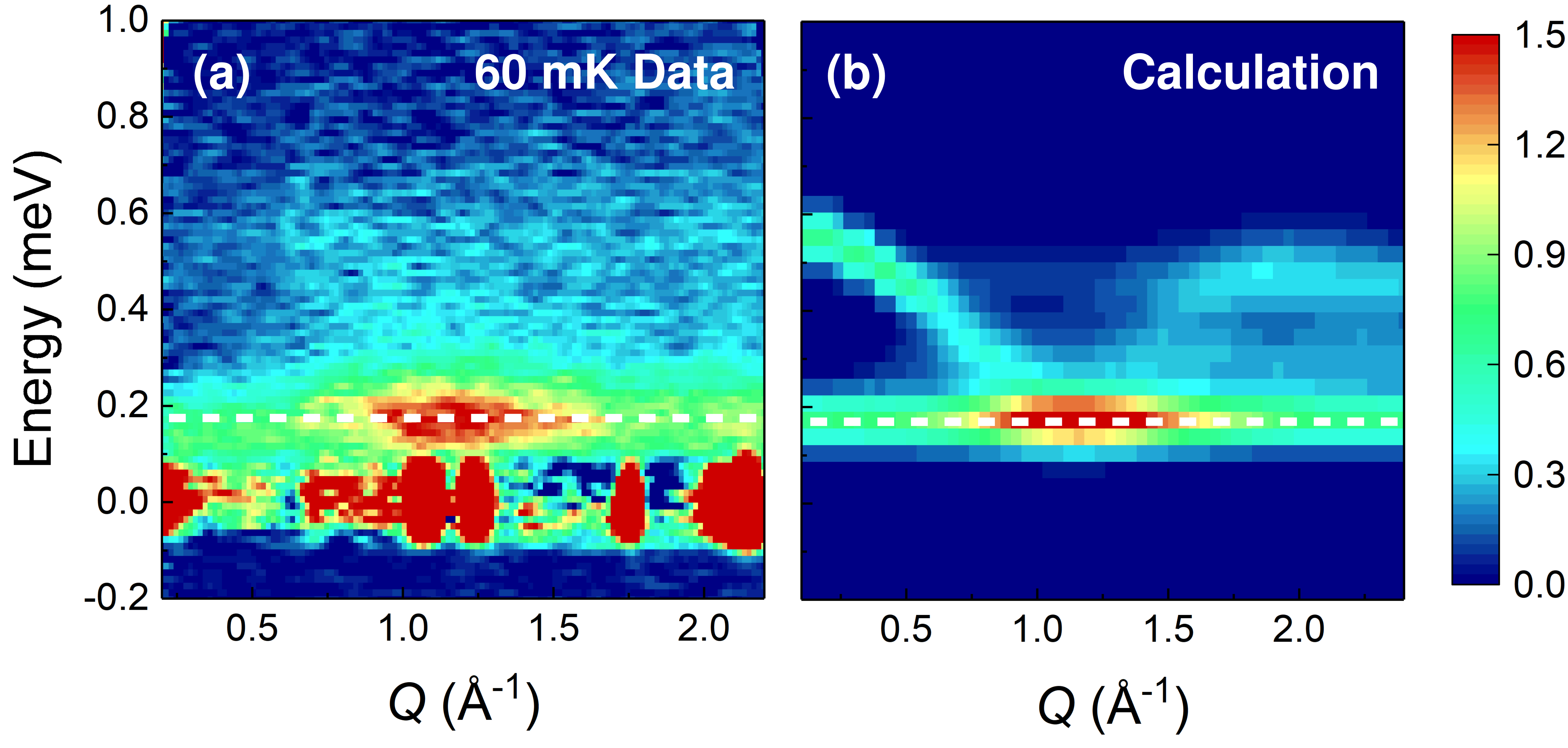}
\par
\caption{Comparison of the (a) measured spin wave spectrum of Er$_2$Pt$_2$O$_7$ at 60~mK with the (b) calculated spin wave spectrum with $J_1=0.10\pm0.05$, $J_2=0.20\pm0.05$, $J_3=-0.10\pm0.03$, and $J_4 =0$~meV. The calculation captures the lower flat band, indicated by the white dashed line, but predicts a dispersive higher energy mode absent in the measurement.}
\label{Calculations}
\end{figure}

We have performed classical spin wave calculations to further investigate the spin excitations of Er$_2$Pt$_2$O$_7$. The powder averaged spin wave spectra were calculated using the anisotropic exchange Hamiltonian~\cite{ross2011quantum,yan2017theory} and further details can be found in the Supplemental Material. We used the experimentally derived exchange parameters for Er$_2$Ti$_2$O$_7$ as an approximate starting point: $J_1=0.10$, $J_2=-0.06$, $J_3=-0.10$, and $J_4 =0$~meV~\cite{savary2012order}. From these values, we carried out a least squares refinement and the best agreement with our experimental spectra for Er$_2$Pt$_2$O$_7$ occurs with $J_1=0.10\pm0.05$, $J_2=0.20\pm0.05$, $J_3=-0.10\pm0.03$, and $J_4 =0$~meV. The calculated spin wave spectrum for this set of parameters is presented in Figure~\ref{Calculations}, where it is presented side-by-side with the lowest temperature experimental data set. This calculated spectra provides a very good description of the low energy flat band, as can be further appreciated by the integrations presented in Figures~\ref{Cuts}(b) and (c). However, there is a striking discrepancy at higher energies: the computed spin excitation spectrum contains an intense, dispersive mode that is not observed in the experimental data (Fig.~\ref{Calculations}(a) and \ref{Cuts}(c)). It is important to emphasize that this intense, dispersive upper band is present in the computed spectra for the entire range of exchange parameters considered in our study. As our exchange parameters for Er$_2$Pt$_2$O$_7$ place it relatively close to the phase boundary between $\Gamma_5$ and $\Gamma_7$, it is possible that enhanced quantum fluctuations due to phase competition are responsible for the breakdown of the quasiparticles associated with this spin wave branch. Similar phenomenology has recently been investigated in Yb$_2$Ti$_2$O$_7$~\cite{thompson2017quasiparticle}. Fruitful comparisons can also be made with Gd$_2$Sn$_2$O$_7$, which also possesses a Palmer-Chalker ground state below $T_N = 1$~K~\cite{wills2006magnetic} but with Heisenberg spins rather than XY anisotropy. In addition to a sharp flat band at low energy, the spin wave spectrum of Gd$_2$Sn$_2$O$_7$ contains at least two additional sharp branches at higher energies~\cite{stewart2008collective}. Thus the breakdown of this upper spin wave branch is not a generic attribute of Palmer-Chalker magnets, evidencing that Er$_2$Pt$_2$O$_7$ experiences stronger quantum effects.

\begin{figure}[tbp]
\linespread{1}
\par
\includegraphics[width=3.3in]{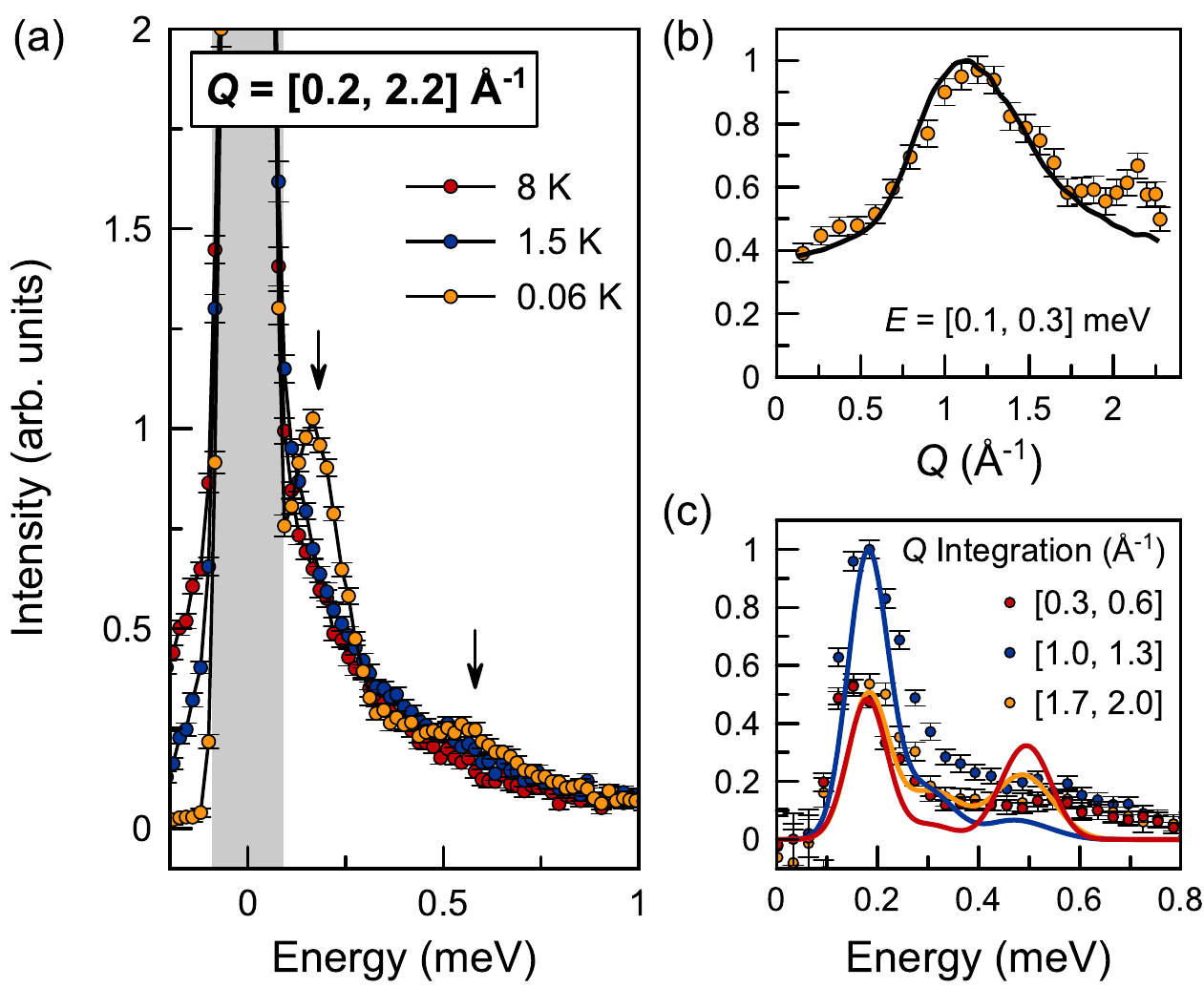}
\par
\caption{Integrated scattered intensity of Er$_2$Pt$_2$O$_7$ as a function of energy transfer over the the full measured $Q$ range, from 0.2~\AA$^{-1}$ to 2.2~\AA$^{-1}$. The gray shaded region indicates the elastic resolution. Below $T_N$, at 0.06~K, the spin excitations are gapped by $0.18\pm0.02$~meV and a weak second band is observed at 0.6~meV. (b) The $Q$ dependence of the lower flat band at $T=0.06$~K, showing good agreement with the spin wave calculation given by the solid line. (c) Integrations over several representative $Q$ intervals. The spin wave calculation provides a good fit to the flat lower band but predicts a second intense branch not observed in the experimental data.}
\label{Cuts}
\end{figure}

The ensemble of Er$_2$Pt$_2$O$_7$'s ground state magnetic properties are remarkable, given that it is structurally so similar to Er$_2$Ti$_2$O$_7$. Indeed the lattice parameters of these two sister compounds differ by less than 0.5\%, far smaller than the 2\% difference with the third sister, Er$_2$Ge$_2$O$_7$, whose magnetic properties are largely unchanged from Er$_2$Ti$_2$O$_7$~\cite{li2014long,dun2015antiferromagnetic}. Comparing Er$_2$Pt$_2$O$_7$ and Er$_2$Ti$_2$O$_7$, we find its N\'{e}el ordering temperature is reduced by a factor of three, from 1.2~K to 0.38~K, and the ordered state itself is altered from $\Gamma_5$ to $\Gamma_7$. Moreover, the spin wave gap of $0.18\pm0.02$~meV is more than triple the $0.053 \pm 0.006$~meV order-by-disorder spin wave gap observed in Er$_2$Ti$_2$O$_7$~\cite{ross2014order}. Despite the lower band being very narrow, the full bandwidth of the inelastic scattering in Er$_2$Pt$_2$O$_7$, 0.6~meV, is still considerably larger than that of Er$_2$Ti$_2$O$_7$, 0.4~meV~\cite{ruff2008spin}. Such observations eliminate simple energetic arguments for Er$_2$Pt$_2$O$_7$'s anomalously low $T_N$. Two considerations are important to understand these surprising differences: (i) The partially occupied platinum $5d$ orbital can facilitate superexchange pathways that are inaccessible in closed shell titanium~\cite{cai2016high,hallas2016relief}, and (ii) The XY pyrochlores live in a rich phase space where modest changes in anisotropic exchange parameters can have a large effect on ground state selection~\cite{yan2017theory}. Indeed, this paradigm predicts that proximity to competing classical phases should manifest as a suppressed ordering temperature~\cite{yan2017theory}. Thus, our observations strongly implicate that Er$_2$Pt$_2$O$_7$ resides in a region of exchange parameter space where it is subject to strong $\Gamma_5$-$\Gamma_7$ phase competition. 

Related phenomenology has previously been observed in the ytterbium family of pyrochlores, including both broad and sharp specific heat anomalies, where the spin dynamics develop well-above $T_N$ or $T_C$~\cite{ross2009two,dun2013yb,robert2015spin,gaudet2016gapless,hallas2016universal}. However, for the ytterbium pyrochlores, this competition is between the ferromagnetic $\Gamma_9$ state and the antiferromagnetic $\Gamma_5$ state~\cite{robert2015spin,jaubert2015multiphase,yan2017theory}. In the case of Er$_2$Pt$_2$O$_7$, it is two antiferromagnetic states, $\Gamma_5$ and $\Gamma_7$, that compete. Thus, we interpret the short-range order at $T^*$ as originating in the spins fluctuating between these two XY states, without breaking the continuous U(1) degeneracy. Then at a lower temperature, $T_N$, a single manifold is uniquely selected. Conversely, no such broad anomaly or unusual spin dynamics are observed in Er$_2$Ti$_2$O$_7$~\cite{ruff2008spin}, which orders at a much higher temperature and for which phase competition is certainly less important~\cite{yan2017theory}.

We have shown that Er$_2$Pt$_2$O$_7$ is the first realization of a Palmer-Chalker ($\Gamma_7$) ground state amongst the XY pyrochlores, with $T_N=0.38$~K and an ordered moment of 3.4(2)~$\mu_{\text{B}}$. The spin dynamics develop well-above the ordering temperature, near $T^*=1.5$~K, the origin of the broad specific anomaly. The dramatically suppressed ordering temperature and change of ground state in Er$_2$Pt$_2$O$_7$ can be understood in the context of strong phase competition. Multiphase competition is already understood to be important within the ytterbium family of pyrochlores and our work shows that this premise can equally be expanded into the erbium pyrochlores. 

\emph{Note added:} Following the submission of this paper, a related manuscript on another erbium XY pyrochlore, Er$_2$Sn$_2$O$_7$, appeared on the arXiv~\cite{petit2017long}. This material is also found to possess a Palmer-Chalker ground state with $T_N=0.1$~K. Evidence is also found for frustration induced by phase competition, consistent with our arguments on its relevance to erbium pyrochlores.

\begin{acknowledgments}
We greatly appreciate the technical support from Juscelino Leao and Yegor Vekhov at the NIST Center for Neutron Research. A.M.H. acknowledges support from the Vanier Canada Graduate Scholarship Program and thanks the National Institute for Materials Science (NIMS) for their hospitality and support through the NIMS Internship Program. This work was supported by the Natural Sciences and Engineering Research Council of Canada. We acknowledge the support of the National Institute of Standards and Technology, U.S. Department of Commerce, in providing the neutron research facilities used in this work. Identification of commercial equipment does not imply endorsement by NIST. C.R.W. acknowledges CFI and the CRC program (Tier II).
\end{acknowledgments}

\bibliography{ErPtO_Ref}

\newpage

\section{SUPPLEMENTAL MATERIAL:}
\section{Experimental Methods}
Er$_2$Pt$_2$O$_7$ can be synthesized in the cubic $Fd\bar{3}m$ pyrochlore structure, in powder form only, using high-pressure techniques. Unlike the rare earth germanium and lead pyrochlores, it is not the relative size of the $A$ and $B$ site cations that necessitates the use of high-pressure synthesis \cite{wiebe2015frustration,hallas2015magnetic}. In fact, the ionic radii of platinum is intermediate to titanium and tin, and is thus well within the ambient pressure stability range for the formation of a pyrochlore phase with many rare earths. Rather, it is the low decomposition temperature of PtO$_2$ which necessitates the use of high-pressure. Stoichiometric quantities of Er$_2$O$_3$ and PtO$_2$ were reacted in gold or platinum capsules at 6 GPa and 1200$^\circ$C for two hours, then rapidly quenched to room temperature before releasing the pressure. Small quantities of unreacted Er$_2$O$_3$ and Pt-metal were removed by repeated washings in aqua regia. Each batch of Er$_2$Pt$_2$O$_7$ was x-rayed to verify phase purity and the total resultant sample was 1.2~grams. The room temperature lattice parameter ($a = 10.13$~\AA) and Curie-Weiss temperature ($\theta_{\text{CW}} = -24.7$~K~\footnote{The Curie-Weiss temperature was obtained by fitting $\chi^{-1}$ between 50~K and 300~K.}) for our sample are consistent with previous reports~\cite{hoekstra1968synthesis,sleight1968new,cai2016high}.\\

Elastic neutron scattering measurements were carried out on the cold neutron triple-axis spectrometer SPINS at the National Institute for Standards and Technology's Center for Neutron Research (NCNR)~\footnote{The collimation on SPINS was guide-80'-80'-open with Be filters both before and after the sample}. These measurements were performed with a monochromatic neutron beam of wavelength 4.04~\AA, giving an energy resolution of $0.34$~meV at the elastic line. Inelastic neutron scattering measurements were performed on the time-of-flight Disc Chopper Spectrometer (DCS) at the NCNR~\cite{copley2003disk}. An incident neutron beam of 5~\AA~was used, corresponding to a maximum energy transfer of 3.3~meV, with an energy resolution of 0.09~meV. Both experiments were performed with an ICE dilution insert in a standard orange ILL cryostat over a temperature range of 60 mK to 8 K. To facilitate thermal equilibration at low temperatures, the sample was wrapped in a piece of copper foil and sealed in a copper sample can with 10 atmospheres of helium. All error bars correspond to one standard deviation. The DCS data were reduced and visualized using the DAVE software package~\cite{azuah2009dave}. The magnetic symmetry analysis was performed with SARAh~\cite{wills2000new} and Rietveld refinements were carried out using FullProf~\cite{rodriguez1993recent}.

\section{Order Parameter}

We used the SPINS spectrometer to measure the scattered elastic neutron intensity at the center of the (111) and (002) Bragg peaks, the results of which are shown in Figure~2(c) of the main manuscript. All data were collected on warming from base temperature, allowing two minutes to equilibrate at each temperature. Our order parameter indicates the formation of a statically ordered state below $T_N=0.38$~K, consistent with the heat capacity measured in Ref.~\cite{cai2016high}. In a narrow temperature range below $T_N$, the data is well fit to a power law dependence,
\begin{equation}
I \propto \left( \frac{T_N - T}{T_N} \right)^{2\beta}
\end{equation}
where $\beta$ is the critical exponent. A fit between 0.38~K and 0.32~K, given by the yellow curve in Figure~2(c) gives $\beta = 0.35 \pm 0.03$, consistent with conventional 3D XY universality~\cite{campostrini2001critical}, as well as the $\beta$ extracted for Er$_2$Ti$_2$O$_7$~\cite{champion2003er}. However, over a wider temperature range, the form of this order parameter is somewhat unusual for its lack of curvature.

\section{Spin Wave Calculation}

\begin{figure*}[tbp]
\linespread{1}
\par
\includegraphics[width=7in]{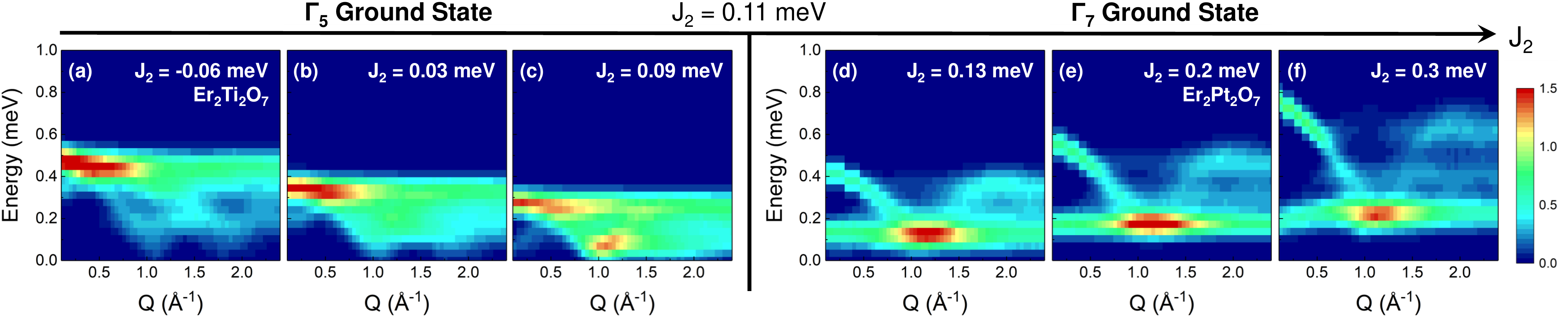}
\par
\caption{Evolution of the powder averaged spin excitation spectrum as a function of $J_2$ where $J_1=0.11$~meV, $J_3=-0.10$~meV, and $J_4=0$~meV. A phase transition between $\Gamma_5$ and $\Gamma_7$ occurs when $J_1=J_2$. The set of exchange parameters displayed in (a) are those that describe Er$_2$Ti$_2$O$_7$. The best qualitative agreement for Er$_2$Pt$_2$O$_7$ is shown in (e), in close proximity to the $\Gamma_5$/$\Gamma_7$ phase boundary.}
\label{Calculations}
\end{figure*}

As described in the text, we fit the exchange parameters of Er$_2$Pt$_2$O$_7$ using an anisotropic exchange Hamiltonian. The nearest-neighbor anisotropic exchange Hamiltonian has been extensively employed to describe the magnetism in other pyrochlore magnets~\cite{ross2011quantum,savary2012order,yan2017theory} and is defined as: 
\begin{eqnarray}
\mathcal{H}_{ex}= \sum_{\langle ij \rangle} J_{ij}^{\mu\nu}S_{i}^{\mu}S_{j}^{\nu}.
\label{eq: HEX}
\end{eqnarray}
The sum $\langle ij \rangle$ runs over all nearest-neighbor bonds. $S_i$ is the magnetic moment operator of an Er$^{3+}$ atom located at site $i$ and is composed of three components: $S_x$, $S_y$ and $S_z$. Due to the symmetry of the pyrochlore lattice, the exchange coupling matrix, $J_{ij}^{\mu\nu}$, can be reduced to four independent exchange parameters~\cite{curnoe2007quantum}. The exchange parameters are defined as $J_1$ (XY term), $J_2$ (Ising term), $J_3$ (symmetric off-diagonal term) and $J_4$ (Dzyaloshinskii-Moriya term).\\

To fit the exchange parameters of Er$_2$Pt$_2$O$_7$, we used the base temperature, $T=0.06$~K, data set shown in Figure 4(a) of the main manuscript, where an empty can data set has been subtracted. The fitted data have also been corrected for the absorption of Er$^{3+}$ in cylindrical geometry. We parameterized the exchange couplings by fitting the scattered intensity as a function of energy transfer for three different integrations in momentum transfer, $Q$. The intervals in momentum transfer, $Q$, used for these three integrations are 0.3 to 0.6~\AA$^{-1}$, 1.0 to 1.3~\AA$^{-1}$ and 1.7 to 2.0~\AA$^{-1}$, as shown in Figure~5(c) of the main text. We further constrained the spin wave calculation by fitting  an energy integration from 0.1 to 0.3~meV, which corresponds to an integration over the flat mode. The resulting integration is shown in Figure 5(b) of the main text and shows the intensity of the flat mode as a function of the momentum transfer, $Q$. Note that the small peak near 2.1~{\AA$^{-1}$} corresponds to crossing the phonon branch extending out of (222), which is the most intense nuclear Bragg reflection. \\  

The linear spin wave calculation was performed using the program SpinW~\cite{toth2015linear} with the anisotropic exchange Hamiltonian of Equation 2 in the Supplemental Material. The exchange parameters $J_1$, $J_2$, and $J_3$ were freely varied, but $J_4$ were kept fixed to 0 as this term has been found to be negligible for all other known XY pyrochlore magnets~\cite{ross2011quantum,savary2012order,guitteny2013palmer}. The best agreement with our measured data was obtained with $J_1=0.10\pm0.05$, $J_2=0.20\pm0.05$, $J_3=-0.10\pm0.03$. The calculated spin wave spectra for this set of exchange parameters is shown in Figure 4(b) of the main text and quantitative comparison is shown in Figure 5(b,c). The energy and momentum transfer dependence of the flat mode is well captured by our model. However, the high energy spin wave branches predicted by the spin wave calculation are not observed in our base temperature inelastic neutron scattering data set. The most striking difference occurs at low $Q$ where a relatively intense and dispersive spin wave branch is expected to be present, but is not visible in our data. Instead, a broad and weak continuum of scattering is observed in higher energy. \\

It is interesting to note that the experimentally determined exchange parameters for Er$_2$Ti$_2$O$_7$ differ from those of Er$_2$Pt$_2$O$_7$ only in their $J_2$ value. The computed spin wave spectra for these two materials are shown in Figure~\ref{Calculations}(a) and \ref{Calculations}(e), respectively. In the nearest-neighbor anisotropic exchange phase space, the transition from $\Gamma_5$ to $\Gamma_7$ occurs when $J_2$ becomes larger than $J_1$. Thus, in the remaining panels of Figure~\ref{Calculations} we have simulated the spin excitation spectrum as a function of increasing $J_2$ keeping all other exchange parameters fixed at the values determined for Er$_2$Ti$_2$O$_7$ and Er$_2$Pt$_2$O$_7$. The relatively flat lower band appears over the entire Palmer-Chalker phase space with dispersive higher energy modes of increasing band width.

\end{document}